\documentstyle[pre,preprint,eqsecnum,aps]{revtex}
\tighten
\begin{document}
\draft
\title{Equations of state for nonlinear sigma-models II:
Relations between resummation schemes, and crossover phenomena}
\author{S.Q.Yang and D.Belitz}
\address{Department of Physics and Materials Science Institute,\\
University of Oregon,\\
Eugene, OR 97403}

\date{\today}
\maketitle

\begin{abstract}
It is shown how a recent method to systematically extrapolate and resum
the loop expansion for nonlinear sigma-models is related to solutions
of the renormalization group equation. This relation is used to
generalize the explicit equations of state obtained previously
to models which display crossover phenomena. As an example we discuss Wegner's
localization model and consider the crossover from symplectic to unitary
symmetry.
\end{abstract}
\narrowtext
\section{INTRODUCTION}
\label{sec:I}

Nonlinear $\sigma$-models have proven useful in Statistical Mechanics
for the description of both ferromagnetism
\onlinecite{BrezinNelson} and Anderson localization of
noninteracting electrons\cite{Wegner79,ZJ}. The description by a nonlinear
$\sigma$-model allows to study these phase transitions
in the vicinity of the lower critical dimension, which is $d_c^- =2$
in either case. For the localization problem this description is
particularly valuable, since no upper critical dimension is known for
the localization, or Anderson, transition,
no mean-field theory exists, and hence
the usual methods of expanding about mean-field theory and the upper
critical dimension are not applicable \cite{LeeRama}. The question then arises
whether it is possible to derive a simple
equation of state \onlinecite{eqstatefootnote} for the
localization problem, which is capable of qualitatively describing
both phases and the transition between them.
The answer turned out to
be affirmative, but in a somewhat restricted sense. An approximate
equation of state was derived for the case of localization in the presence
of time reversal invariance \cite{selfcons}. It turned out to correctly
reproduce the critical behavior known from renormalization group (RG)
methods to first order in an $\epsilon$-expansion about $d=2$ \cite{Wegner79}.
This equation of state became known as the 'self-consistent theory
of Anderson localization', and it has been derived by a variety of
techniques \cite{selfcons,Hikami}.
However, it was not obvious how to generalize this
theory to models with broken time reversal invariance,
or how to systematically include
higher orders in the $\epsilon$-expansion. Attempts to improve on
this situation by exponentiating perturbation theory gave results
that again were satisfactory only in the case of time reversal
invariance\cite{PW}.

In a recent paper\cite{I}, to be denoted by I, the present authors have
derived equations of state for nonlinear $\sigma$-models that are
based on a systematic loop expansion, and give physical results for
localization models both with and without time reversal invariance.
The method was also applied to nonlinear $\sigma$-models for
Heisenberg ferromagnets.
The basic idea in I was to obtain the RG $\beta$-function to a given order
in the loop expansion, truncate it, and then construct a perturbation
series for the relevant physical quantities (e.g. the conductivity
in the case of localization, or the magnetization in the
case of magnets) that yields the truncated
$\beta$-function exactly. The resummation of this extrapolated
perturbation series led to transcendental equations whose solution
gave the desired equation of state.

Such an extrapolation and resummation of perturbation theory is
essentially what the RG equation also achieves\cite{ZJ}.
Indeed, one of the derivations of the 'self-consistent localization
theory' made use of the RG equation\cite{Hikami}. It is therefore
natural to ask whether there is a simple relation between these two
methods. In the present paper we show that a simple relation
does indeed exist, and how the RG equation method can be used to
derive equations of state identical to those obtained in
Ref. \onlinecite{I}. We then use this
insight to derive an explicit equation of
state for the case of localization with weakly broken time reversal
invariance, when there is a crossover between two fixed points governed
by different symmetries. The same method can be used to describe
anisotropic Heisenberg ferromagnets\cite{thesis}.
However, the localization case
is slightly more difficult since there the two fixed points governed by
different symmetries appear at different orders in the loop expansion.
Here we restrict ourselves to the localization problem, and just
mention that essentially the same discussion, with some simplifications,
applies to anisotropic magnets.

\section{THE MODEL, AND ITS RENORMALIZATION}
\label{sec:II}

\subsection{Definition of the Model}
\label{subsec:II.A}

As in I, we consider Wegner's model for the localization of noninteracting
electrons \cite{Wegner79}.
Both the model and our notation will be the same as in I, and so
we will be very brief. Wegner's model is a nonlinear $\sigma$-model
defined on the symmetric
spaces $Sp(2N)/Sp(N)\times Sp(N)$ and $U(2N)/U(N)\times U(N)$ for systems
with and without time reversal invariance, respectively. Quenched
disorder is handled in these models by considering $N$ replicas of
the system, and taking the limit $N\rightarrow 0$ ("replica limit")
after calculations. The action reads
\begin{mathletters}
\label{eqs:2.1}
\begin{equation}
S = \frac{1}{T} \int d{\bf x}\ tr\,\left(\nabla Q({\bf x})\right)^2
    + \frac{H}{T} \int d{\bf x}\ tr\left[\Lambda Q({\bf x})\right]\quad,
\label{eq:2.1a}
\end{equation}
where $Q$ is a $2N\times 2N$ matrix subject to the constraints,
\begin{equation}
Q^2({\bf x}) = \openone_{2N}\quad,\quad tr\ Q({\bf x}) = 0\quad,
\label{eq:2.1b}
\end{equation}
and $\Lambda$ is a block-diagonal matrix,
\begin{equation}
\Lambda = \left(\matrix{\openone_N & 0\cr
                            0      & -\openone_N\cr}\right)\quad,
\label{eq:2.1c}
\end{equation}
\end{mathletters}%
with $\openone_N$ the $N\times N$ unit matrix. The symmetry of the action is
symplectic or unitary, respectively, depending on
whether the matrix elements of $Q$ are quaternion or complex valued.
We will refer to these models as the symplectic and the unitary model,
respectively.
In both cases the matrix $Q$ is hermitian, and in the
symplectic model it is subject to the additional constraint
$Q=C^T Q^T C$, where $C = i\,\sigma_x\otimes\sigma_y$ with Pauli matrices
$\sigma_{x,y}$.

It is well known that each of the two models defined by Eqs.\ (\ref{eqs:2.1})
describes a localization transition of
noninteracting electrons \cite{LeeRama,R}. The bare
coupling constant $T$ in Eq.\ (\ref{eq:2.1a}) is proportional to the
conductivity in self-consistent Born approximation (SCBA);
it is a measure of disorder. $H=\Omega\pi N_F/4$ with $\Omega$ the
external frequency, and $N_F$
the density of states at the Fermi
level. The symplectic model describes the localization transition in
systems that are time reversal invariant. If time reversal invariance
is broken, e.g. by magnetic impurities, the critical behavior is
instead described by the unitary model.

For later reference we also define an additional term \cite{ELK},
which we add to
the action for the symplectic model,
\begin{equation}
\delta S = \frac{G}{4T}\ \int d{\bf x}\ tr\ [Q({\bf x}),-i\sigma_z
                                \otimes\openone_N]^2\quad,
\label{eq:2.2}
\end{equation}
where $[a,b]$ denotes the commutator of $a$ and $b$.
The term $\delta S$ breaks the symmetry of the symplectic model
action, and for any nonzero $G$ the asymptotic critical behavior of
the phase transition described by the symplectic model plus the term
$\delta S$ is the same as that of the unitary model. For small values
of $G$, however, the asymptotic critical behavior is preceded by a
region that displays a crossover between the symplectic and the unitary
fixed point. This model is relevant for localization with weakly broken
time reversal invariance, e.g. in the presence of a small concentration
of magnetic impurities \cite{ELK}.
It also describes the localization of third
sound waves in superfluid He films on a rough substrate with a superimposed
uniform superfluid flow \cite{CohenMachta}.

It is convenient to eliminate the nonlinear constraint,
Eq.\ (\ref{eq:2.1b}), by parameterizing
the matrix $Q$ as,
\begin{equation}
Q = \left(\matrix{\left[\openone_N - q q^{\dagger}\right]^{1/2}
                             - \openone_N & q\cr
            q^{\dagger} & -\left[\openone_N - q^{\dagger} q\right]^{1/2}
                                         + \openone_N\cr}
    \right)\quad,
\label{eq:2.3}
\end{equation}
and to expand $q$ and $q^{\dagger}$ in a quaternion basis,
\begin{equation}
q = \sum_{r}\ q_r\ \tau_r\quad,\quad
q^{\dagger} = \sum_{r}\ q^{\dagger}_r \tau_r^{\dagger}\quad,
\label{eq:2.4}
\end{equation}
where $r = 0,1,2,3$ and $r=0,3$ for the symplectic and unitary models,
respectively, and $\tau_0 = \openone_2$, $\tau_{1,2,3} = -i\sigma_{x,y,z}$.
The $q_r$ are $N\times N$ matrices.

Let us consider the two-point $q$-vertex function, which is the inverse
of the $q$-$q$ propagator. The most general model is the symplectic one
with the symmetry breaking term, Eq.\ (\ref{eq:2.2}), added. One obtains
two different 2-point vertex functions for the channels $r=0,3$ and $r=1,2$,
respectively. To Gaussian order they read,
\begin{mathletters}
\label{eqs:2.5}
\begin{equation}
\Gamma^{(2)}_{p-h} (p) = \frac{1}{T}\left(p^2 + H\right)
                                             + O\left(T^0\right)\quad,
\label{eq:2.5a}
\end{equation}
\begin{equation}
\Gamma^{(2)}_{p-p} (p) = \frac{1}{T}\left(p^2 + H + G\right)
                                             + O\left(T^0\right)\quad.
\label{eq:2.5b}
\end{equation}
\end{mathletters}%
The subscripts $p-h$ and $p-p$ refer to the particle-hole ($r=0,3$) and
particle-particle ($r=1,2$) channels, respectively \cite{LeeRama,R}.
$G=0$ corresponds to
the symplectic model. In this case both vertex functions coincide
and are soft, or
massless, i.e. they vanish with vanishing wavenumber and frequency. The
soft modes describe diffusion in the particle-hole and particle-particle
channels. A nonzero $G$ breaks time reversal invariance, and
gives the particle-particle channel a mass. The particle-hole channel
remains massless because of particle number conservation.

The most interesting physical quantity in the context of the
localization problem is the
conductivity, $\sigma$. It is related to $\Gamma^{(2)}_{p-h}$ by
\begin{equation}
\sigma = \lim_{p\rightarrow 0}\ \frac{\partial}{\partial p^2}\
                                           T\Gamma^{(2)}_{p-h}(p)\quad,
\label{eq:2.6}
\end{equation}
where we have normalized $\sigma$ by its SCBA value.

\subsection{Renormalization}
\label{subsec:II.B}

It is well known that the symplectic and unitary models defined by Eqs.\
(\ref{eqs:2.1}) are renormalizable with two renormalization constants,
one for the coupling constant $T$, and one field renormalization
constant \cite{ZJ}.
In the replica limit the field renormalization vanishes, and one has
only one nontrivial renormalization constant, which we denote by $Z_t$.
The symmetry breaking term, Eq.\ (\ref{eq:2.2}), requires one additional
renormalization constant, $Z_g$, for the coupling constant $G$. Denoting
the renormalized disorder, frequency, mass, and conductivity by $t$,
$\omega$, $g$, and $\sigma_r$, respectively, the relations between the
bare quantities and their renormalized counterparts are
\begin{mathletters}
\label{eqs:2.7}
\begin{equation}
T = \kappa^{-\epsilon} Z_t t\quad,
\label{eq:2.7a}
\end{equation}
\begin{equation}
\Omega = \kappa^2 Z_t \omega\quad,
\label{eq:2.7b}
\end{equation}
\begin{equation}
G = \kappa^2 Z_g Z_t g\quad,
\label{eq:2.7c}
\end{equation}
\begin{equation}
\sigma = Z_t \sigma_r\quad,
\label{eq:2.7d}
\end{equation}
\end{mathletters}%
with $\epsilon = d-2$, and $\kappa$ an arbitrary RG momentum scale.
{}From the requirement that the bare theory must be independent of
$\kappa$ one obtains a RG equation \onlinecite{ZJ} for $\sigma_r$,
\begin{equation}
\left[\epsilon - \frac{1}{t}\beta(t,g)
               + \beta(t,g)\ \frac{\partial}{\partial t}
               + \gamma(t,g)\frac{\partial}{\partial g}
               + \left(\frac{1}{t}\beta(t,g) - d\right)\omega
                                \frac{\partial}{\partial\omega}\right]
       \sigma_r (t,\omega,g) = 0\quad.
\label{eq:2.8}
\end{equation}
Here $\beta(t,g)=dt/d\ln\kappa$ and $\gamma(t,g)=dt/d\ln\kappa$, with
the derivatives taken at a fixed bare theory, are RG functions. They can
be obtained in a loop expansion from the perturbation expansion for
the vertex function, Eqs.\ (\ref{eqs:2.5}), by standard methods. We
have performed the calculation to two-loop order with the result,
\begin{mathletters}
\label{eqs:2.9}
\begin{equation}
\beta(t,g) = \epsilon t - \frac{t^2}{4}\ \frac{1}{1+g}
                        - \frac{t^3}{32}\ \frac{g}{1+g} + O(t^4)\quad,
\label{eq:2.9a}
\end{equation}
\begin{equation}
\gamma(t,g) = -2g + \frac{t}{4}\ \frac{g^2}{1+g} + O(t^2)\quad.
\label{eq:2.9b}
\end{equation}
\end{mathletters}%
In the limits $g=0$ (symplectic model), and $g=\infty$ (unitary model)
Eq.\ (\ref{eq:2.9a}) reproduces well-known
results \cite{ZJ}. The $\beta$-functions
for these two cases are actually known to five-loop
order \cite{Wegner89,Hikami92}.
However, for the sake of simplicity we will restrict ourselves to
the two-loop result as contained in Eqs.\ (\ref{eqs:2.9}). Consequently,
all critical exponents, scaling functions, etc. will be valid to
two-loop order only.
Inspection of Eqs.\ (\ref{eqs:2.9})
shows that they allow for two nontrivial fixed points (FP), viz.
$(t^*,g^*) = (4\epsilon,0)$, and
$(t^*,g^*) = (\sqrt{32\epsilon},\infty)$. The former is
unstable with respect to $g$, while the latter is stable. These two FP
correspond to the symplectic and unitary universality classes, respectively,
of the Anderson transition.

\section{THE EQUATION OF STATE}
\label{sec:III}

We now obtain the desired equation of state, i.e. the conductivity
as a function of disorder $t$, frequency $\omega$, and symmetry
breaking parameter $g$, from the RG equation, Eq.\ (\ref{eq:2.8}).
For simplicity we do this for the renormalized theory. Since all
bare and renormalized quantities are proportional to each other,
the same functional relations hold for both the bare and the
renormalized theory, and we will drop the subscript on $\sigma_r$.

\subsection{The Symplectic and Unitary Cases Revisited}
\label{subsec:III.A}

Let us first reconsider the symplectic and unitary models. In either
case there is no $g$-dependence, and the $\beta$-function is given
by Eq.\ (\ref{eq:2.9a}) with $g=0$ or $g=\infty$, respectively. At
zero frequency the PDE (\ref{eq:2.8})
(without the $\partial/\partial g$-term) then turns into an ODE, which
is readily solved in terms of a quadrature. With our normalization
condition for the conductivity, $\sigma(t\rightarrow 0,\omega = 0)
\rightarrow 1$ (see Eq.\ (\ref{eq:2.6}),
we obtain for the static conductivity,
$\sigma_0 (t) \equiv \sigma (t,\omega = 0)$,
\begin{equation}
\sigma_0 (t) = \exp \left( \int_0^t d\tau\ \left[\frac{1}{\tau}
               - \frac{\epsilon}{\beta(\tau)}\right]\right)\quad.
\label{eq:3.1}
\end{equation}
Of course, $\sigma_0 (t) =0$ is also a solution of the RG equation.
The physical conductivity is determined by Eq.\ (\ref{eq:3.1}) only
for $t<t_c$, where $t_c$ is given by $\beta(t_c)=0$. For $t>t_c$
the physical conductivity vanishes, and $\sigma_0 (t)$ as given by
Eq.\ (\ref{eq:3.1}), and interpreted as a principal value integral,
has no direct physical meaning. For $t\alt t_c$
Eq.\ (\ref{eq:3.1}) yields,
\begin{mathletters}
\label{eqs:3.2}
\begin{equation}
\sigma_0 (t) = c(t) \vert 1 - t/t_c\vert ^s\quad,
\label{eq:3.2a}
\end{equation}
where the function $c(t)$ is analytic at $t=t_c$, and the conductivity
exponent $s$ is related to the correlation length exponent $\nu$,
\begin{equation}
s = \nu\epsilon = -\epsilon/\beta'(t_c)\quad.
\label{eq:3.2b}
\end{equation}
\end{mathletters}%
The first equality in Eq.\ (\ref{eq:3.2b}) is known as Wegner's scaling
law \cite{Wegner76}.

Using standard methods \onlinecite{Zach}
it is now easy to express the general solution
of the PDE, Eq.\ (\ref{eq:2.8}), for $g=0$ or $g=\infty$ in terms of
$\sigma_0 (t)$. The associated system of ODEs is
\begin{equation}
\frac{d\ t}{\beta(t)} = \frac{d\ \ln\omega}{-d+\beta(t)/t}
                      = \frac{d\ \ln\sigma}{-\epsilon+\beta(t)/t}\quad.
\label{eq:3.3}
\end{equation}
Let us consider the equalities relating the first term and the third
term, and the first term and the second term, respectively. The
general solutions of these two ODEs are,
\begin{mathletters}
\label{eqs:3.4}
\begin{equation}
u_1 = \sigma/\sigma_0(t)\quad,
\label{eq:3.4a}
\end{equation}
\begin{equation}
u_2 = \omega\ t^{-2/\epsilon} [\sigma_0(t)]^{-d/\epsilon}\quad,
\label{eq:3.4b}
\end{equation}
\end{mathletters}%
where $u_1$ and $u_2$ are integration constants. The general solution
of the PDE is $\tilde F(u_1,u_2)=0$, with $\tilde F$ an arbitrary function
of two variables. This can be rewritten as,
\begin{equation}
\sigma(t,\omega) = \sigma_0 (t)\ F\left(\omega\,t^{2/\epsilon}\,
                    \sigma_0^{-d/\epsilon}(t)\right)\quad,
\label{eq:3.5}
\end{equation}
with $F$ an arbitrary function of one variable. In order to determine
$F$ we need to impose a boundary condition. Since the conductivity at
large frequencies ($\omega\approx 1$) is qualitatively insensitive to
the disorder, we can normalize the conductivity by its value at
$\omega = 1$,
\begin{equation}
\sigma(t,\omega =1) = 1\quad.
\label{eq:3.6}
\end{equation}
{}From Eq.\ (\ref{eq:3.5}) this
yields a transcendental equation for $F$, which can be rewritten as
\begin{equation}
x = [t_F(F(x))]^{2/\epsilon}\ [F(x)]^{d/\epsilon}\quad,
\label{eq:3.7}
\end{equation}
where $t_F(x)$ is obtained as the inverse of the relation between $F$
and $\sigma_0$,
\begin{equation}
F = 1/\sigma_0(t)\quad\rightarrow\quad t = t_F(F)\quad.
\label{eq:3.8}
\end{equation}
Note that $t_F(F)$ has two branches. Finally, from Eq.\
(\ref{eq:3.5}) we know that $F(x)=\sigma/\sigma_0(t)$, with the
argument of the function $F$ given by $x=\omega\,t^{2/\epsilon}\,
[\sigma_0(t)]^{-d/\epsilon}$. Using this is Eq.\ (\ref{eq:3.7}) we
obtain the equation of state in the form,
\begin{equation}
\frac{\omega}{\sigma^{d/\epsilon}} = t^{-2/\epsilon}\
  \left[t_F\left(\sigma/\sigma_0(t)\right)\right]^{2/\epsilon}\quad.
\label{eq:3.9}
\end{equation}

Equation (\ref{eq:3.9}) is identical with the result that was obtained
by different methods in I. This shows that the resummation scheme of I
is equivalent to solving the RG equation, Eq.\ (\ref{eq:2.8}), with
the boundary condition given by Eq.\ (\ref{eq:3.6}).
Let us write down explicit results by using the
$\beta$-function, Eq.\ (\ref{eq:2.9a}), to two-loop order. The generalization
to higher loop orders is straightforward, and has been given in I.
To two-loop order we obtain for the symplectic model,
\begin{mathletters}
\label{eqs:3.10}
\begin{equation}
\frac{\omega}{\sigma^{d/\epsilon}} = \left(t/t_c\right)^{-2/\epsilon}\
  \left[1 - \frac{1-t/t_c}{\sigma}\right]^{2/\epsilon}\quad,
\label{eq:3.10a}
\end{equation}
with $t_c=4\epsilon$, and for the unitary one,
\begin{equation}
\frac{\omega}{\sigma^{d/\epsilon}} = \left(t/t_c\right)^{-2/\epsilon}\
  \left[1 - \frac{1-t^2/t_c^2}{\sigma^2}\right]^{1/\epsilon}\quad.
\label{eq:3.10b}
\end{equation}
\end{mathletters}%
with $t_c=\sqrt{32\epsilon}$.
Equation (\ref{eq:3.10a}) is the well-known result of
Ref. \onlinecite{selfcons}. The above
derivation is essentially identical with the one given by
Hikami \cite{Hikami}.
Equation (\ref{eq:3.10b}) is the corresponding result for the unitary
model, which was first derived in I. We note that the derivation
using the RG equation is shorter and more transparent than the
resummation scheme used in I.

\subsection{The Crossover from Symplectic to Unitary Critical Behavior}
\label{subsec:III.B}

We will now apply the same method to construct an equation
of state for our more general model, defined by the symplectic model
with the symmetry breaking term, Eq.\ (\ref{eq:2.2}).
The RG equation is given by Eqs.\ (\ref{eq:2.8}), (\ref{eqs:2.9}).
Let us first consider the zero frequency conductivity,
$\sigma_0(t,g) \equiv \sigma(t,\omega=0,g)$. In this case the
system of associated ODEs consists of two independent equations,
which we choose as,
\begin{mathletters}
\label{eqs:3.11}
\begin{equation}
\frac{d\ t}{\beta(t,g)} = \frac{d\ g}{\gamma(t,g)}\quad,
\label{eq:3.11a}
\end{equation}
\begin{equation}
\frac{d\ \ln\sigma_0}{\beta(t,g)/t - \epsilon} = \frac{d\ g}{\gamma(t,g)}
\label{eq:3.11b}
\end{equation}
\end{mathletters}%
The general solution of Eq.\ (\ref{eq:3.11a}) can be written in the form,
\begin{mathletters}
\label{eqs:3.12}
\begin{equation}
f_1(t,g) = u_1\quad,
\label{eq:3.12a}
\end{equation}
with $u_1$ an integration constant.
The general solution of Eq.\ (\ref{eq:3.11b}) reads,
\begin{equation}
\sigma_0\ \exp\left(\int dg \frac{\epsilon - \beta(t,g)/t}
                                           {\gamma(t,g)}\right) = u_2\quad,
\label{eq:3.12b}
\end{equation}
where $u_2$ is a second integration constant. The general solution of the
PDE for $\sigma_0$ is given by $\tilde F(u_1,u_2)=0$ with $\tilde F$ an
arbitrary function, or
\begin{equation}
\sigma_0(t,g) = f(u_1)\ \tilde\sigma_0(t,g)\quad,
\label{eq:3.12c}
\end{equation}
with $u_1$ a function of $t$ and $g$ according to Eq.\ (\ref{eq:3.12a}),
$f$ an arbitrary function of $u_1$, and
\begin{equation}
\tilde\sigma_0(t,g) =  \exp\left(\int dg
                                      \frac{\beta(t,g)/t - \epsilon}
                                           {\gamma(t,g)}\right)\quad,
\label{eq:3.12d}
\end{equation}
\end{mathletters}%
Note that for determining the indefinite integral in Eqs.\ (\ref{eq:3.12b}),
(\ref{eq:3.12d}), $t$ must be expressed in terms of $g$ and $u_1$ by means
of Eq.\ (\ref{eq:3.12a}). $u_1$ is kept fixed for the integration, and
then replaced by $f_1(t,g)$ again.

Equation (\ref{eq:3.12c}) is the desired equation of state at zero
frequency. What remains to be done is to determine $u_1$ or $f_1$
explicitly as a function of $t$ and $g$, and to
determine the unknown function $f$ by imposing suitable
boundary conditions. This is a substantially more complicated task than
for the pure symplectic or unitary model.
We first derive an explicit expression for the
function $f_1$, Eq.\ (\ref{eq:3.12a}).
Even with the simple two-loop approximations for the RG functions,
Eqs.\ (\ref{eqs:2.9}), we have been unable to find a closed form
representation for this function. However, it can be discussed
analytically in certain limits. Let us first consider the case of
weak symmetry breaking, $g<<1$. For $g=0$ there is a fixed point
already at one-loop order, and we can restrict ourselves to a
one-loop approximation for the $\beta$-function,
$\beta(t,g)\approx \epsilon t - (t^2/4)(1-g)$. Since both $g$ and $t$
are small compared to unity, it is then sufficient to use the zero-loop
approximation for the RG function $\gamma$, $\gamma(t,g)\approx -2g$.
Equation (\ref{eq:3.11a}) is then readily integrated with the result,
\begin{mathletters}
\label{eqs:3.15}
\begin{equation}
f_1(t,g\rightarrow 0) =
            \frac{1}{g^{\epsilon/2}}\left[\frac{1}{t} - \frac{1}{4\epsilon}
                                    + O(g)\right]\quad.
\label{eq:3.15a}
\end{equation}
In the same limit one obtains from Eq.\ (\ref{eq:3.12d})
$\tilde\sigma_0(t,g) = tg^{\epsilon/2}$. With $t_c^0\equiv t_c(g=0) =
4\epsilon$ the critical disorder for the symplectic
model, the zero frequency conductivity can be written,
\begin{equation}
\sigma_0(t,g\rightarrow 0) = tg^{\epsilon/2}\ f\left(
     \frac{1-t/t_c^0}{tg^{\epsilon/2}}\right)\quad.
\label{eq:3.15b}
\end{equation}
\end{mathletters}
For $t$ near $t_c^0$ we thus obtain the scaling form which was first
derived by Khmel'nitskii and Larkin from a scaling
hypothesis \cite{LarkinKhmel}. We see
that both the value of the conductivity at $t=t_c^0$ and the shift of
the mobility edge are proportional to $g^{1/\phi}$ with $\phi=2/\epsilon
+O(1)$ the crossover exponent\cite{LarkinKhmel,Opper}.
Comparing Eqs.\ (\ref{eq:3.15b}) and
(\ref{eq:3.10a}) we also see that for large arguments (i.e., for
$g\rightarrow 0$), the function $f$ behaves like $f(x\rightarrow\infty)
\sim x$.

Similarly, one obtains in the limit of large $g$,
\begin{mathletters}
\begin{equation}
f_1(t,g\rightarrow\infty) =
             \frac{1}{g^{\epsilon}}\left[\frac{1}{t^2} - \frac{1}{32\epsilon}
                      + \frac{1}{32g}\right]
         \equiv\frac{1}{g^{\epsilon}}\left[\frac{1}{t^2} - \frac{1}{t_c^2(g)}
                                           \right]\quad.
\label{eq:3.16a}
\end{equation}
For $\tilde\sigma_0$ one obtains again
$\tilde\sigma_0(t,g) = tg^{\epsilon/2}$, so we have for the conductivity,
\begin{equation}
\sigma_0(t,g\rightarrow\infty) = tg^{\epsilon/2}\ f\left(
     \frac{1-t^2/t_c^2(g)}{t^2g^{\epsilon}}\right)\quad.
\label{eq:3.16b}
\end{equation}
Close to criticality this leads to the scaling form
\begin{equation}
\sigma_0 = \sqrt{1-t^2/t_c^2(g)}\
                \Phi\left(\frac{1/g}{[1-t^2/t_c^2(g)]^{\phi'}}\right)\quad,
\label{eq:3.16c}
\end{equation}
\end{mathletters}%
with $\phi'=-1/\epsilon$. Here we have used that $f(x\rightarrow 0)
\sim \sqrt{x}$, which can be seen by comparing Eqs.\ (\ref{eq:3.16b})
and (\ref{eq:3.10b}). Notice the qualitative difference between
the scaling forms at small $g$, Eq.\ (\ref{eq:3.15b}), and at large
$g$, Eq.\ (\ref{eq:3.16c}). In the former case, the conductivity is
driven to zero because of a zero in the scaling function $f$, and the
asymptotic critical behavior is determined by the behavior of $f$
in the vicinity of that zero. This gives rise to a crossover from
critical behavior governed by the symplectic fixed point to the
asymptotic behavior which is governed by the unitary fixed point.
We will see this crossover behavior in more detail below, when we determine
the function $f$ explicitly. In the latter case
the conductivity vanishes due to the
prefactor in Eq.\ (\ref{eq:3.16c}). The function $\Phi$ merely
provides corrections to this asymptotic scaling behavior, and there
is no crossover. The exponent $\phi'$ is a corrections to scaling
exponent. The reason for this difference is that $g$ is a relevant
operator with respect to the symplectic fixed point, while $1/g$
is an irrelevant operator with respect to the unitary one. It is
well known that a relevant operator leads to a crossover, while
irrelevant ones lead to corrections to scaling. Our RG treatment
thus does not confirm a previous conjecture that the scaling form
of the conductivity is the same for both small and large $g$ \cite{Opper}.

We now turn to the task of determining the scaling function $f_1$
for all $g$. A problem one encounters in that context is that
the solution of the equation $f_1(t,g)=0$, which determines the
critical disorder of the system as a function of $g$, develops
from a single root, $t=t_c^0$ for $g=0$, into a double root,
$t=\pm t_c^{\infty}$ for $g=\infty$. This problem can be solved as
follows. It is convenient to switch from the variables $(t,g)$ to
new variables $(x,y)=(1/t,g/(1+g))$. Equation (\ref{eq:3.11a})
reads in terms of these variables,
\begin{equation}
\frac{dx}{dy} = \frac{-\epsilon x + (1-y)/4 + y/32x}
                     {-2y(1-y) + y^2(1-y)/4x}\quad.
\label{eq:3.17}
\end{equation}
Let us now consider flow equations for $x$ and $y$ that lead to
the same $f_1$, determined by Eq.\ (\ref{eq:3.17}), as the
original ones, but are obtained by multiplying numerator and
denominator on the r.h.s. of Eq. (\ref{eq:3.17}), by $-x$,
\begin{mathletters}
\label{eqs:3.18}
\begin{equation}
\frac{dx}{dl} = \epsilon x^2 - x(1-y)/4 - y/32\quad,
\label{eq:3.18a}
\end{equation}
\begin{equation}
\frac{dy}{dl} = 2xy(1-y) - y^2(1-y)/4\quad.
\label{eq:3.18b}
\end{equation}
\end{mathletters}%
These flow equations allow for four fixed points, viz.
$(x_1,y_1)=(1/4\epsilon,0)$, $(x_2,y_2)=(1/\sqrt{32\epsilon},1)$,
$(x_3,y_3)=(-1/\sqrt{32\epsilon},1)$, and $(x_4,y_4)=(0,0)$, and
the flow diagram is shown in Fig.\ 1. Note that only the region
$x>0,y>0$ in this diagram is physical, and that the region $x<0$
serves a purely mathematical purpose. Important trajectories in
this flow diagram are the curves $x=\chi_+(y)$, and $x=\chi_-(y)$, which
connect the fixed points (1) and (2), and (3) and (4), respectively.
We have determined only the qualitative behavior of these curves
to the extent that we will need them, viz.
\begin{mathletters}
\label{eqs:3.19}
\begin{equation}
\chi_+(y\rightarrow 0) = x_1-{\rm const}\times y^{1/\phi}\quad,
\label{eq:3.19a}
\end{equation}
\begin{equation}
\chi_+(y\rightarrow 1) = x_2 + \frac{1}{8}\left(1 + O(\sqrt{\epsilon})\right)
                                             (1-y)\quad,
\label{eq:3.19b}
\end{equation}
and
\begin{equation}
\chi_-(y\rightarrow 0) = 0\quad,
\label{eq:3.19c}
\end{equation}
\begin{equation}
\chi_-(y\rightarrow 1) = -x_2 + \frac{1}{8}\left(1 + O(\sqrt{\epsilon})\right)
                                             (1-y)\quad.
\label{eq:3.19d}
\end{equation}
\end{mathletters}%
The constant in Eq.\ \ref{eq:3.19a} can only be determined numerically,
and of $\chi_-(y\rightarrow 0)$ we will need to know only that it vanishes.
We now write $f_1$ as a function of $x$ and $y$ in the following form,
\begin{equation}
f_1(x,y) = [x-\chi_+(y)]\ [x-\chi_-(y)]\ \frac{y^{-\epsilon/2}(1-y)^{\epsilon}}
                                        {y + x(1+y)^{\epsilon/2}}\
                                   G(x,y)\quad.
\label{eq:3.20}
\end{equation}
The explicit part of this equation contains all qualitative features of
the flow. In particular, all of the singular behavior has been built in,
as well as the known behavior at small and large $g$, respectively. The
function $G(x,y)$ is therefore a harmless function of both of its variables,
and we approximate $G(x,y)\approx 1$. Changing the variables back to
$t$ and $g$, we then have
\begin{mathletters}
\label{eqs:3.21}
\begin{equation}
f_1(t,g) = [1/t-1/t^*(g)]\ [1/t-1/t_c(g)]\
           \left[\frac{g}{1+g} +
             \frac{1}{t}\ \left(\frac{1}{1+g}\right)^{\epsilon/2}\right]^{-1}\
           g^{-\epsilon/2}\ (1+g)^{-\epsilon/2}\quad,
\label{eq:3.21a}
\end{equation}
where
\begin{equation}
t_c(g\rightarrow 0) = 4\epsilon + {\rm const}\times g^{\epsilon/2}\quad,
\label{eq:3.21b}
\end{equation}
\begin{equation}
t_c(g\rightarrow\infty) = \sqrt{32\epsilon} - {\rm const}/g\quad,
\label{eq:3.21c}
\end{equation}
is the critical disorder as a function of $g$, and
\begin{equation}
t^*(g\rightarrow 0) \rightarrow \infty\quad,
\label{eq:3.21d}
\end{equation}
\begin{equation}
t^*(g\rightarrow\infty) = -\sqrt{32\epsilon} - {\rm const}/g\quad.
\label{eq:3.21e}
\end{equation}
\end{mathletters}%
is the auxiliary curve corresponding to $\chi_-(y)$.

Now we are in a position to construct an approximate equation of state
for the case of zero frequency that contains all qualitative features
which we have discussed so far. Let us return to Eq.\ (\ref{eq:3.12c}).
As we have seen, $\tilde\sigma_0$ is equal to $tg^{\epsilon/2}$ both
for small $g$ and for large $g$, and we approximately put $\tilde\sigma_0 =
tg^{\epsilon/2}$ everywhere. From Eqs.\ (\ref{eq:3.15b}) and
(\ref{eq:3.16b}) we see that the function $f(u_1)$ is proportional
to $u_1$ for large $u_1$  and proportional to
$\sqrt{u_1}$ for small $u_1$. This function can therefore be adequately
modeled by $f(x)=\sqrt{x(1+x)}$. We now have,
\begin{equation}
\sigma_0(t,g) = tg^{\epsilon/2}\ \left[f_1(t,g)(1+f_1(t,g))\right]^{1/2}\quad,
\label{eq:3.22}
\end{equation}
with $f_1(t,g)$ from Eqs.\ (\ref{eqs:3.21}).

Equations (\ref{eqs:3.21}), (\ref{eq:3.22}) are the desired result for
the equation of state, i.e. for the conductivity as a function of the
disorder $t$ and the symmetry breaking parameter $g$ at zero frequency.
Let us briefly discuss this result. For $g\rightarrow 0$,
$1/t^*$ vanishes, $t_c$ approaches $4\epsilon$, $f_1$ diverges like
$g^{-\epsilon/2}$,
and $\sigma_0$ becomes independent of $g$. Equation (\ref{eq:3.22}) then
reproduces the result for the symplectic model, Eq.\ (\ref{eq:3.10a})
at $\omega=0$. For $g\rightarrow\infty$, $1/t^*$ approaches
$-1/t_c(g=\infty)=-1/\sqrt{32\epsilon}$, and $f_1$ vanishes like
$g^{-\epsilon}$. Again $\sigma_0$ becomes independent of $g$, and
Eq.\ (\ref{eq:3.22}) reproduces the result for the unitary model,
Eq.\ (\ref{eq:3.10b}) at $\omega=0$. If one approaches the critical
disorder, $t\rightarrow t_c(g)$, at fixed but small $g$ ($g<<1$), then
the critical region consists of two regimes. For $\vert t-t_c(g)\vert
>> g^{\epsilon/2}$ one has $f_1>>1$, and $\sigma_0$ shows a behavior
characteristic of approaching the symplectic localization transition.
In particular, the critical exponent $s=1$. However, when
$\vert t-t_c(g)\vert << g^{\epsilon/2}$ one has $f_1<<1$, and the
asymptotic approach to the phase transition is governed by the
critical behavior of the unitary model. This is the crossover behavior
induced by the instability of the symplectic fixed point with respect
to the relevant operator $g$. With increasing $g$ the crossover region
moves away from the critical point, and for sufficiently large values
of $g$ it is outside of the critical region. The critical behavior is
then entirely that of the unitary model, with corrections to scaling
due to the irrelevant variable $1/g$.

We finally take the frequency into account.
We choose the third ODE associated with the PDE, Eq.\ (\ref{eq:2.8}), as,
\begin{mathletters}
\begin{equation}
\frac{d\ln\omega}{-d+\beta(t,g)/t} = \frac{dg}{\gamma(t,g)}\quad.
\label{eq:3.13a}
\end{equation}
The general solution reads,
\begin{equation}
\omega t^{2/\epsilon}[\tilde\sigma_0(t,g)]^{-d/\epsilon} = u_3\quad,
\label{eq:3.13b}
\end{equation}
\end{mathletters}%
with $u_3$ a third integration constant, in addition to $u_1$ and $u_2$.
The general solution of the full
PDE can then be written as,
\begin{equation}
\sigma(t,\omega,g) = \tilde\sigma_0(t,g)\
 G\left(f_1(t,g),\omega t^{2/\epsilon}\
                             [\tilde\sigma_0(t,g)]^{-d/\epsilon}\right)\quad,
\label{eq:3.14}
\end{equation}
with $G$ an arbitrary function of two variables. The
function $f$ of Eq.\ (\ref{eq:3.12c}) is related to $G$ by $G(x,0)=f(x)$.
We now impose
the same boundary consition as in Sec.\ \ref{subsec:III.A}, namely
$\sigma(t,\omega=1,g)=1$. The transcendental equation for $G(u_1,x)$
analogous to Eq.\ (\ref{eq:3.7}) then reads,
\begin{equation}
x = \left[t_G\left(G(u_1,x),u_1\right)\right]^{2/\epsilon}
    \left[G(u_1,x)\right]^{d/\epsilon}\quad,
\label{eq:3.23}
\end{equation}
Here $t_G$, which is a function of two variables, is obtained as follows.
In the relation $G=1/\tilde\sigma_0(t,g)$ we express $g=g(t,u_1)$ as a
function of $t$ and $u_1$ according to Eq.\ (\ref{eq:3.12a}). $t_G$ is
then given explicitly as a function of $G$ and $u_1$,
\begin{equation}
G = 1/\tilde\sigma_0\left(t,g(t,u_1)\right)\qquad\rightarrow\qquad
        t = t_G(G,u_1)\quad.
\label{eq:3.24}
\end{equation}
Now we restore the frequency dependence. With
$x=\omega t^{2/\epsilon} [\tilde\sigma_0(t,g)]^{-d/\epsilon}$ one
has $G(u_1,x) = \sigma/\tilde\sigma_0$. Inserted into
Eq.\ (\ref{eq:3.23}) this yields,
\begin{equation}
\frac{\omega}{\sigma^{d/\epsilon}} = t^{-2/\epsilon}\
      t_G\left(\sigma/\tilde\sigma_0(t,g),f_1(t,g)\right)\quad.
\label{eq:3.25}
\end{equation}
This is the desired equation of state for the general model. It
contains the previous results for the symplectic ($g=0$) and
unitary ($g=\infty$) models as special cases, and explicitly
describes the crossover as demonstrated above. It also gives
a physically correct description of both the metallic and the
insulating phases. This has been discussed in detail in I,
and can be seen from Eqs.\ (\ref{eqs:3.10}), in the limits
$g=0$ and $g=\infty$. Since the frequency dependence of the
conducticity is qualitatively the same in both limits, this is
also true for arbitrary values of $g$, and hence for Eq.\ (\ref{eq:3.25}).

\acknowledgments

We thank Ted Kirkpatrick for helpful discussions.
This work was supported by the NSF under grant number
DMR-92-09879.

\begin{figure}
\caption{Schematic flow diagram for the flow equations, Eqs.\
 (\protect\ref{eqs:3.18})}
\label{fig:1}
\end{figure}

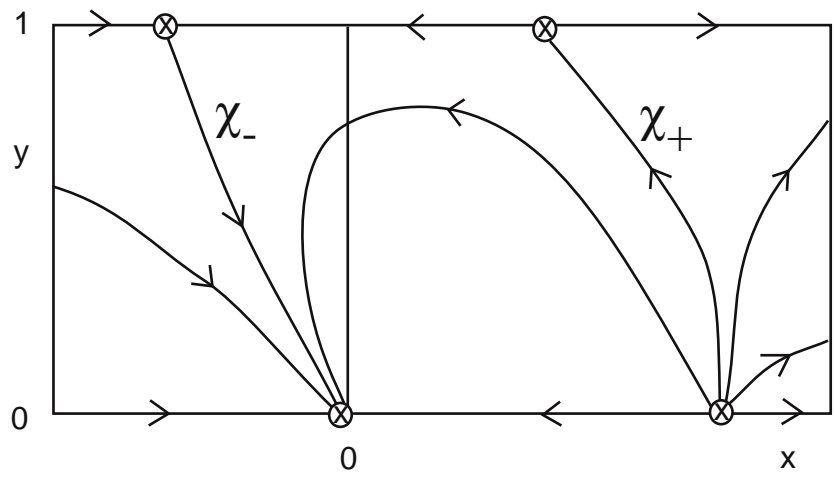
\end{document}